\documentclass[twocolumn]{IEEEtran}
\usepackage{graphicx,subfigure}
\usepackage{enumerate}
\usepackage{color}
\usepackage{pifont}
\usepackage{float}
\usepackage{epsfig,url}
\usepackage[makeroom]{cancel}
\usepackage{multirow}
\usepackage{pbox}
\usepackage{lipsum}
\usepackage{arydshln}
\usepackage{cite,amssymb,amsthm}
\usepackage{tablefootnote}
\usepackage{amsmath}
\usepackage{arydshln}
\usepackage[ruled, vlined]{algorithm2e}
\usepackage{epstopdf}
\usepackage{tikz}
\usetikzlibrary{calc, shapes, arrows.meta}

\newcommand{\mup}{\mu \text{PMU}}
\newcommand{\bs}{\boldsymbol} 

\newcommand{\mb}{\mathbf}
\usepackage{color}
\newtheorem*{definition}{Definition}
\begin{document}

\title{Low-Resolution Fault Localization Using Phasor Measurement Units with Community Detection}
%On Identifiability of Fault Location in Power Distribution Grids
%\author{Mahdi Jamei, Anna Scaglione, Sean Peisert}
\author{
    \IEEEauthorblockN{Mahdi Jamei\IEEEauthorrefmark{1}, Anna Scaglione\IEEEauthorrefmark{1}, Sean Peisert\IEEEauthorrefmark{2}}\\
    \IEEEauthorblockA{\IEEEauthorrefmark{1}Arizona State University,
     \{mjamei, ascaglio\}@asu.edu}\\
    \IEEEauthorblockA{\IEEEauthorrefmark{2}Lawrence Berkeley National Laboratory,
    sppeisert@lbl.com} 
\\[-3.0cm]
 \thanks{
    This research was supported in part by the Director, Cybersecurity, Energy Security, and Emergency Response, Cybersecurity for Energy Delivery Systems program, of the U.S. Department of Energy, under contract DE-AC02-05CH11231 and DE-OE0000780.  Any opinions, findings, conclusions, or recommendations expressed in this material are those of the authors and do not necessarily reflect those of the sponsors of this work.}
}
\maketitle
\begin{abstract}
A significant portion of the literature on fault localization assumes (more or less explicitly) that there are sufficient reliable measurements to guarantee that the system is observable. While several
heuristics exist to break the observability barrier, they mostly rely on recognizing spatio-temporal patterns, without giving insights on how the performance are tied with the system features and the sensor deployment. 
In this paper, we try to fill this gap and investigate the limitations and performance limits of fault localization using Phasor Measurement Units (PMUs), in the low measurements regime, i.e., when the system is unobservable with the measurements available. Our main contribution is to show how one can leverage the scarce measurements to localize different type of distribution line faults (three-phase, single-phase to ground, ...) at the level of sub-graph, rather than with the resolution of a line. 
%Specifically, we show that, depending on the deployment and the number of sensors on the grid, communities of neighboring nodes emerge in which any line fault is in roughly equal agreement with the data, as long as a fault is in this subset of lines; lines in other communities, instead, create a larger residual. These limitations mainly come from the nature of the bus impedance matrix, inevitable approximations of the fault current and lack of enough sensors compared to the size of the grid.  The take away message is that, while it is hard to determine where exactly the fault occurred without full observability, 
%the fault can be localized up to the resolution of these communities. 
We show that the resolution we obtain is strongly tied with the graph clustering notion in network science.     
\end{abstract}
\IEEEpeerreviewmaketitle
\vspace{-.5cm}
\section{Introduction}\label{sec.intro}
%New methods to improve Distribution Management Systems (DMSs) are actively researched right now, particularly as it pertains to data analytics. The operators need to make sense of the information gathered from the field to gain situational awareness, and to be able to operate the grid in an efficient and reliable manner. 
Phasor Measurement Units' (PMU) data have the ability to provide much more accurate event detection capabilities, even with relatively few measurements. 
Ardekanian et al.,~\cite{8273895}, for example, use PMU data to detect and localize a change in the admittance matrix of a distribution network. Zhou et al.,~\cite{zhou2017partial} empoly PMUs to detect events in the distribution grid when only partial information is available. Our previous work~\cite{jamei2017anomaly} proposes a hierarchical architecture for event detection in distribution system using PMU data when only very few sensors are available. Except for Ardekanian et al's work~\cite{8273895}, the literature on event detection is often in a {\it low measurements regime}, that arises when the number of measurements is very small compared to system size, so the system is unobservable.  
%which makes the problem significantly under-determined and therefore extremely sensitive to noise. In other words, when $K \ll M$ we should expect that the fault localization is very imprecise.

Event detection is, however, insufficient in most cases. The localization of a line fault is a classic problem in power systems management. In fact, it is an essential part of any event detection scheme, since the operators need to locate the faulty section for isolation and service restoration. 
What we refer to as line fault is the short-circuit of a single-phase, two phase, or three-phase of a line with each other and/or with ground with or without a fault resistance. As a result, a large magnitude of fault current is withdrawn from  the sources to provide current for the short-circuited location.  

Both for transmission and distribution grids, 
fault detection and localization, particularly using PMU data, is still a very active area of research, which focuses more broadly on understanding the root cause of abnormal changes recorded in the PMU measurements. Most of the work in the {\it low measurement regime} is in the distribution section of the grid.
%However, this regime is important in general when considering adversarial communications and false data injection attacks in particular~\cite{kim2011strategic}. In fact, the ability of using a reduced subset of sensors to gain information on what is plausible in areas that are not directly observed can be used as a tool to reveal stealth data attacks, since all subsets selected will have to agree on a common version of the facts regarding the presence or absence of faults in a certain area of the system. 
For example, Zhu et al.,~\cite{zhu1997automated} propose an automated fault localization and diagnosis for radial networks: specifically, using measurements from the substation, the algorithm first finds a set of plausible fault locations and then run a diagnosis to rank the different possibilities. Min and Santoso~\cite{min2017dc} propose a technique to remove the DC offset in the phasor data and improve the algorithm intended to locate a momentary fault. 
%Focusing on the communication network, Kahyap et al., \cite{kashyap2015automated} implement a fault location and isolation algorithm in a distributed fashion.  
Lee~ \cite{lee2014automatic} uses synchronized voltage phasor measurements to search for a fault in a radial network in a timely manner. Dzafic et al.,~\cite{dvzafic2018fault} propose a graph marking approach to spot the location of a fault. 
There are a number of non-parametric methods that exploit spatio-temporal patterns, as well. Specifically,  
Jian et al.,~\cite{jiang2014fault} extract the time-frequency signatures of voltage and frequency from a dictionary using {\it matching pursuit}~\cite{mallat1993matching}, followed with a clustering algorithm for fault detection. Using wavelet analysis on the voltage waveform generated during a fault-induced transients, Borghetti et al.,~\cite{borghetti2008continuous} obtain the location of a fault in the distribution network.
While the exploitation of temporal patterns helps in the localization, they do not provide an understanding on how the performance is affected by the grid parameters and the sensor deployment.

{\it Contribution:} To dig deeper in the low measurement regime, in this work we do not look at temporal features  (which can always be incorporated in the algorithm) and take inspiration from Brahma's work~\cite{brahma2011fault} to construct our models. Brahma proposes a method using the bus impedance matrix of the systems and pre/post fault voltage and current measurements to pinpoint the location of a fault. 
The contribution of this paper comes from unveiling the specific structure of the errors that localization  algorithm based on PMU data tend to make, showing the fact that the errors swap nodes within very specific sub-graphs of the original grid topology. We can consider these sub-graphs as \textit{communities}, in which nodes are clustered and 
show that {\it community level fault localization} is possible, even when the measurements are too few to have an accurate answer. 
%This in turn means that the algorithm still provides a useful outcome if it is interpreted as the localization of the fault within a certain community, accepting an answer which has lower resolution regarding the event that occurred. Our work introduces a new graph clustering approach for power systems, in the context of fault localization capabilities using PMU sensors in the low measurements regime.  We show that {\it community level fault localization} is possible, even when the measurements are too few to have an accurate answer. 
To the best of our knowledge, the connection between the resolution of fault localization in power systems and graph clustering is new. However, we acknowledge the graph clustering work in the transmission grid (see e.g., \cite{wang2010compressing}, \cite{sanchez2014hierarchical}), in a different context (not for fault localization). 
%and follow up work \cite{cuffe2017visualizing,raak2016data},
\section{Single Fault Localization}\label{sec.main_formulation}
Since distribution lines are untransposed, and because of the existence of single phase and two phase laterals, we prefer to use a formulation that explicitly includes the phase-domain (and not the sequence-domain) voltage and current. For a network of size $N$ the nodal voltages and injection currents vectors are denoted by:
\begin{align*}
\mb{V}=\big[V_1, V_2, \hdots, V_N\big]^T,~~~
\mb{I}=\big[I_1, I_2, \hdots, I_N\big]^T
\end{align*}     
where depending on the number of phases connected to node $i$, $V_i$ and $I_i$ can be row vectors of size 1, 2 or 3; the resulting $\mb{V}$ and $\mb{I}$ are $M\times 1$. It is well-known that the nodal voltages and injection currents satisfy Ohm's law:
\begin{align}
\mb{V}=\mb{Z}\mb{I},
\label{eq.ohm_law}
\end{align}
In \eqref{eq.ohm_law}, all the sources are modeled with their Norton equivalent and therefore their internal impedance is also included in the bus impedance matrix $\mb{Z}$.    
Suppose a fault occurs at bus $j$ and let $\mb{V}_0$ and $\mb{I}_0$ denote the pre-fault  and $\mb{V}_F$ and $\mb{I}_F$ denote the post-fault nodal voltages and currents. Using \eqref{eq.ohm_law} the following two relationships hold:
\begin{align}
\label{eq.pre}
\mb{V}_0=\mb{Z}\mb{I}_0,~~
\mb{V}_F=\mb{Z}(\mb{I}_F+\mb{I}_E)
%\label{eq.post}
\end{align} 
where $\mb{I}_E$ is a sparse vector with non-zero elements at locations corresponding to faulty phases of bus $j$, modeling the current injected by the fault at bus $j$ (negative of the withdrawn current).  
Subtracting the pre-fault from the post-fault voltages:
\begin{align}
\underbrace{\mb{V}_F-\mb{V}_0}_{\bs{\delta}\mb{V}}=\mb{Z}(\underbrace{\mb{I}_F-\mb{I}_0}_{\bs{\delta}\mb{I}}+\mb{I}_E)
\label{eq.post_pre}
\end{align} 
%We also define $\mb{I}_G$ to be the sub-vector containing the non-zero elements of the vector $\mb{I}_E$.
Let $K$ denote the total number of phases that are monitored by the PMUs in the grid. For example, if we have two PMUs, where one is connected to a three phase node and the other is connected to a single phase node, $K=3+1=4$. Let:
\begin{align} 
\bs{\Pi}=\big(\bs{\Pi}^T_a~|~\bs{\Pi}^T_u\big)^T \in \{0,1\}^{M \times M}
\end{align}
that parses the voltage and current measurements into available and unavailable measurements, where $\bs{\Pi}_a \in \{0,1\}^{K \times M}$ picks the available measurements and $\bs{\Pi}_u \in \{0,1\}^{(M-K) \times M}$ selects the unavailable ones. Pre-multiplying both sides of \eqref{eq.post_pre} by $\bs{\Pi}_a$ and also replacing $\mb{Z}$ in the first term with $\mb{Z}\bs{\Pi}^{-1}\bs{\Pi}$ and rearranging some terms, one can write: 
\begin{align}
\begin{split}
\overbrace{\bs{\Pi}_a\bs{\delta}\mb{V}}^{\bs{\delta}\mb{V}_a}&=\overbrace{\big(\bs{\Pi}_a\mb{Z}\bs{\Pi}^{-1}\big)}^{\mb{Z}_a}(\bs{\Pi}~\bs{\delta}\mb{I}+\bs{\Pi}~\mb{I}_E)\\
\bs{\delta}\mb{V}_a&=\big(\mb{Z}_{aa}~|~\mb{Z}_{au}\big)
\begin{pmatrix}
\bs{\Pi}_a\bs{\delta}\mb{I}\\ \bs{\Pi}_u\bs{\delta}\mb{I}
\end{pmatrix}+\mb{Z}_a\mb{\tilde{I}}_E\\
&=\big(\mb{Z}_{aa}~|~\mb{Z}_{au}\big)
\begin{pmatrix}
\bs{\delta}\mb{I}_a\\ \bs{\delta}\mb{I}_u
\end{pmatrix}+\mb{Z}_a\mb{\tilde{I}}_E\\
&=\mb{Z}_{aa}\bs{\delta}\mb{I}_a+\mb{Z}_{au}\bs{\delta}\mb{I}_u+\mb{Z}_a\mb{\tilde{I}}_E
\label{eq.available_voltage}
\end{split}
\end{align}
where $\mb{Z}_{aa}$ and $\mb{Z}_{au}$ are the blocks of impedance matrix connecting nodes with available data to the available and unavailable ones, respectively. Also, $\mb{\tilde{I}}_E$ is the reordered vector $\mb{{I}}_E$ with indices corresponding to the nodes with available data in the top part and the rest at the bottom part.  

Brahma~\cite{brahma2011fault} proposes the formulation above, assuming that the available measurements come from the sensors placed at the head of the substation and next to each distributed generator.  
Then, assuming that the term $\mb{Z}_{au}\bs{\delta}\mb{I}_u$ in \eqref{eq.available_voltage} is small, the location of the faulty bus in his work can be found by solving the following least-square problem~\cite{brahma2011fault}:
\begin{align}
\ell^*=\underset{\ell \in \mathcal{F}(t)}{\text{argmin }}||\bs{\delta}\mb{V}_a-\mb{Z}_{aa}\bs{\delta}\mb{I}_a-\mb{Z}_a\mb{\tilde{I}}_{E,\ell}||^2_2
\label{eq.single_form1}
\end{align}
where $\mathcal{F}(t)$ is the union of candidate fault locations subsets for fault type $t$.
%, where each subset contains node indices corresponding to the faulty phases of the faulty bus. For example, if there is a three-phase fault, a member of $\mathcal{F}(t)$ is itself a set with three members corresponding to the three phase node indices in the vector $\mb{\tilde{I}}_E$ of a three-phase bus, where a fault could potentially occur. 
$\mb{\tilde{I}}_{E,\ell}$ is a vector whose non zero entries correspond to a certain fault location $\ell$ and $\mb{I}_G$ is a vector containing non-zero entries of $\mb{\tilde{I}}_{E,\ell}$. 
Note that in \eqref{eq.single_form1}, the entries of $\mb{I}_G$ are not directly measured. To address this problem, Brahma~\cite{brahma2011fault} proposes to approximate the vector $\mb{I}_G$ by adding up the current injected by each source to the grid corresponding to the faulty phases\footnote{The vector $\mb{I}_G$ is of size 1, 2 or 3 depending on how many phases are impacted by the fault.} and subtracting the current that sources have been providing for the loads in the pre-fault condition. This puts a requirement on the $\mup$ placement strategy since each source requires a $\mup$, to which it is connected.
%\footnote{We wish to remark that if a source is connected to a delta-grounded wye transformer that is usually the type of transformer that can be found in a substation, the sensor should be placed on the grounded-wye side since the zero sequence information of the fault current is unobservable to the sensor on the delta side and therefore the approximation of the fault is not accurate.}

Before going through further discussions about the performance analysis of single fault localization, we propose the following modification to the method in \eqref{eq.single_form1}.  
In order to make the term $\mb{Z}_{au}\bs{\delta}\mb{I}_u$ as small as possible, the constant impedance loads, constant power loads, and capacitors/reactors are also included in the $\mb{Z}$ matrix. This modeling is accurate for constant impedance loads. However, for constant power loads their equivalent impedance at the nominal voltage is included in the bus impedance matrix, and the deviation of the actual consumed power from the nominal is included in the nodal injection vector. This modeling implies that the vectors $\mb{I}_F$ and $\mb{I}_0$ are small and, accordingly, $\bs{\delta}\mb{I}_u$ is small as well. From now on, we assume that this modification is included in the method when we refer to \eqref{eq.single_form1}.

\section{Fault Localization Perfomeance Analysis}\label{sec.identify}
As stated before, the method in \ref{eq.single_form1} is plagued by ambiguities in locating the fault precisely when $K \ll M$. The first source of ambiguity comes from the fact that the vector $\mb{I}_G$ is an approximation of the fault current. Therefore, a fault might be mis-located if the approximated fault current is close to the actual fault current if there was a fault at the mis-located bus. The other source of ambiguity appears if the columns in the $\mb{Z}_a$ for two locations are very similar; in fact, $\mb{Z}_a$ is a fat matrix, so it is inevitable that columns will be correlated.

\begin{figure}[ht]
\centering
\includegraphics[trim={0 0.5cm 0 0.75cm},width=0.36\linewidth]{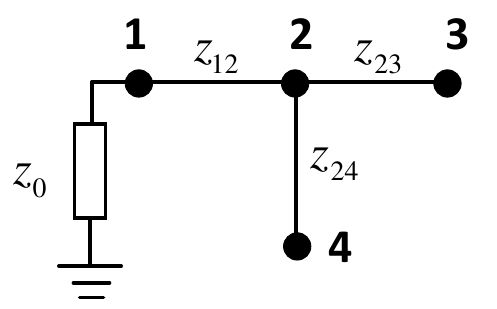}
\caption{One-Line Diagram of an Example Radial Network}
\label{fig.illustration}
\end{figure}

To illustrate why the latter occurs, consider the one-line diagram given in Fig.~\ref{fig.illustration}, where bus 1 is assumed to be the source bus and is modeled with the Norton equivalent.\footnote{The network is modeled with its positive sequence impedance values in this example for simplicity of illustration.}
The bus impedance matrix of this network is given below, where the columns are ordered based on the bus numbers.
\begin{align}
\hspace{-0.25 cm}\mb{Z}=
\begin{pmatrix}
z_0&z_0&z_0&z_0\\
z_0&z_0+z_{12} & z_0+z_{12}&    z_0+z_{12}\\
z_0&z_0+z_{12} &z_0+z_{12}+z_{23} &  z_0+z_{12}   \\
z_0& z_0+z_{12} &  z_0+z_{12}&  z_0+z_{12}+z_{24}   \\
\end{pmatrix}
\end{align}
Now assume that we have sensors at bus 1 and 2, so that:
$$\mb{Z}_a=\begin{pmatrix}
z_0&z_0&z_0&z_0\\
z_0&z_0+z_{12} & z_0+z_{12}&    z_0+z_{12}\\
\end{pmatrix}
$$ 
and assume that we have access to the exact value of the fault current. Based on the given criterion for fault detection in \eqref{eq.single_form1}, the norm in the objective function would be the same if the fault happened at bus 2, 3 or 4, since the columns are exactly the same. This situation can actually happen in radial grids, if one looks at the way the bus impedance matrix is constructed. To provide some intuition, consider a radial grid with a single source and negligible line shunt components. Since there is no loop, the bus impedance matrix can be constructed starting from the source bus and adding new nodes. When a new node $j$ is added to an existing node $i$, the column and row corresponding to the existing node is copied on the column and row corresponding to the new node and the entry $[\mb{Z}]_{jj}=[\mb{Z}]_{ii}+z_{\text{added line}}$. Thus, the entries of column $i$ and $j$ are very similar. This process is sketched below, when a new node (Node 3) is added to the existing node (Node 2).
\begin{figure}[ht]
\centering
\includegraphics[trim={0 0 0 0.75cm},width=0.36\textwidth]{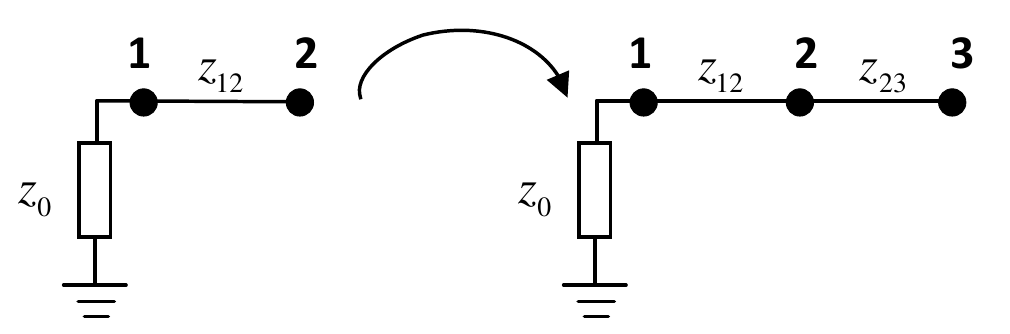}
\label{fig.illustration_NewNode}
\vspace{-0.25cm}
\end{figure}
\begin{align*}
\mb{Z}=\hspace{-0.1cm}
\begin{pmatrix}
z_0&z_0\\
z_0&\tikz[remember picture] \coordinate (oval3); z_0+ \tikz[remember picture] \coordinate (oval1); z_{12}
\end{pmatrix}
\hspace{-0.1cm}\rightarrow \hspace{-0.1cm}
\mb{Z}=\hspace{-0.1cm}
\begin{pmatrix}
z_0&z_0&z_0\\
z_0&z_0+z_{12}&z_0+\tikz[remember picture] \coordinate (oval2);z_{12}\\
z_0&\tikz[remember picture] \coordinate (oval4);z_0+z_{12}&z_0+\tikz[remember picture] \coordinate (oval5);z_{12}+z_{23}
\end{pmatrix}
\begin{tikzpicture}[overlay,remember picture]
\node[ellipse, minimum width=1.4cm, minimum height=1.25cm,yshift=0.25cm, xshift=-0.2cm, draw, dashed](c1) at (oval1) {};
\node[ellipse, minimum width=1.5cm, minimum height=0.95cm,yshift=0.25cm, xshift=-0.15cm, draw, dashed](c2) at (oval2) {};
\path (c1.north east) edge[-{Latex}, bend left=20] (c2);
\node[ellipse, minimum width=2.2cm, minimum height=0.5cm,yshift=0.1cm, xshift=0.2cm, draw, dashed](c3) at (oval3) {};
\node[ellipse, minimum width=2.2cm, minimum height=0.5cm,yshift=0.05cm, xshift=0.35cm, draw, dashed](c4) at (oval4) {};
\path (c3.south west) edge[-{Latex}, bend left=-10] (c4.west);
\node[ellipse, minimum width=2.25cm, minimum height=0.52cm,yshift=0.05cm, xshift=0.35cm, draw, red](c5) at (oval5) {};
\end{tikzpicture}
\end{align*}
In fact, even if the columns are not exactly the same, they are very similar and therefore correlated. Hence, an estimation error can easily occur, due to the noise in measurements or poor approximation of the fault current, and lead to the selection of an incorrect location. An example of this is when there are sensors at buses 2 and 3, and the impedance of line 2-3 is small.
$$
\mb{Z}_a=\begin{pmatrix}
z_0&z_0+z_{12} & z_0+z_{12}&    z_0+z_{12}\\
z_0&z_0+z_{12} &z_0+z_{12}+z_{23} &  z_0+z_{12} 
\end{pmatrix}
$$  
In this case, distinguishing a fault between buses 2, 3 and 4 would not be easy. 
This observation conforms to the topological proximity of the nodes in a radial network, where neighboring nodes tend to contribute to similar columns in the bus impedance matrix.  
\subsection{Communities of Neighboring Nodes}\label{sec.communities}
The discussion above indicates that 
the errors that algorithm tends to make are swaps of nodes within very specific sub-graphs of the original grid topology, which have associated columns in the bus impedance matrix that are very similar. We call these sub-graphs, {\it communities}. Clearly, these communities are also dependent on the location of the sensors. Before digging into this phenomenon further, we notice the term $\mb{Z}_{au}\bs{\delta}\mb{I}_u$ in \eqref{eq.single_form1} can be viewed as a colored noise. Whitening the noise is appropriate to ensure that it does not line up in preferential directions. To appreciate the effect that the whitening has, in the following we will express the whitened data model directly through the bus admittance matrix. Since the whitened model has the same sparsity as that of the graph topology, this will aid our analysis on how the graph structures contributes to the clusters of similar columns in the sensing matrix (the matrix that is pre-multiplied by $\mb{\tilde{I}}_{E}$). 
We first need to find the relationship between blocks of bus impedance matrix and bus admittance matrix corresponding to observable and unobservable nodes. 
\begin{align}
{\begin{pmatrix}
\mb{Y}_{aa}&\mb{Y}_{au}\\
\mb{Y}^T_{au}&\mb{Y}_{uu}
\end{pmatrix}}^{-1}=
\begin{pmatrix}
\mb{Z}_{aa}&\mb{Z}_{au}\\
\mb{Z}^T_{au}&\mb{Z}_{uu}
\end{pmatrix}
\end{align}  
It is known from algebra that for block matrix of the form above, where $\mb{Y}_{aa}$ and $\mb{Y}_{uu}-\mb{Y}^T_{au}\mb{Y}_{aa}^{-1}\mb{Y}_{au}$ are invertible matrices, one can write~\cite{bernstein2005matrix}:
\begin{align}
\label{eq.Zaa}
\mb{Z}_{aa}&=(\mb{Y}_{aa}-\mb{Y}_{au}\mb{Y}_{uu}^{-1}\mb{Y}^T_{au})^{-1} \\
\mb{Z}_{au}&=-\mb{Z}_{aa}\mb{Y}_{au}\mb{Y}^{-1}_{uu}
\label{eq.Zau}
\end{align} 
Using \eqref{eq.Zaa} and \eqref{eq.Zau}, we can represent the blocks of $\mb{Z}_a$ in terms of blocks of $\mb{Y}$.
Pre-multiplying both sides of \eqref{eq.available_voltage} by $\mb{Z}^{-1}_{aa}$ using the definition of $\mb{Z}_{aa}$ in \eqref{eq.Zaa}, we can write:
\begin{align}
\mb{b}-(\mathcal{I}~|~-\mb{Y}_{au}\mb{Y}^{-1}_{uu})\mb{\tilde{I}}_E=-\mb{Y}_{au}\mb{Y}^{-1}_{uu}\bs{\delta}\mb{I}_u
\label{eq.Yform_1}
\end{align} 
where $\mb{b}$ is defined as follows:
\begin{align*}
\mb{b}=\mb{Z}^{-1}_{aa}\bs{\delta}\mb{V}_a-\bs{\delta}\mb{I}_a.
\end{align*}
Let the singular value decomposition of $\mb{Y}_{au}\mb{Y}^{-1}_{uu}$ to be:
\begin{align}
\mb{Y}_{au}\mb{Y}^{-1}_{uu}=\mb{U}\mb{S}\mb{W}^H
\end{align}
where $\mb{U}$ is $K\times \tilde{K}$ and $\tilde{K} \leq K$,  $\mb{S}$ is a diagonal matrix containing the $\tilde{K}$ all the non-zero singular values of $\mb{Y}_{au}\mb{Y}^{-1}_{uu}$  and $\mb{W}^H$ is of size $\tilde{K} \times (M-K)$. 
%Let $\mb{S}$ denote a square and diagonal matrix containing the non-zero block of $\mb{S}$. Clearly, if $\mb{Y}_{au}\mb{Y}^{-1}_{uu}$ is of full row rank, $\mb{S}$ would be of size $K \times K$. Also let $\mb{U}$ and $\hat{\mb{V}}$ denote the left and right singular vectors corresponding to the non-zero singular values. Then we can write:
%\begin{align}
%\mb{Y}_{au}\mb{Y}^{-1}_{uu}=\mb{U}\mb{S}\mb{W}^H
%\end{align}   
To whiten the noise in \eqref{eq.Yform_1}, we pre-multiply both sides of \eqref{eq.Yform_1} by $\mb{R}=\mb{S}^{-1}\mb{U}^H$ to obtain: 
\begin{align}
\mb{R}\mb{b}-(\mb{S}^{-1}\mb{U}^H~|~-\mb{W}^H)\mb{\tilde{I}}_E=-\mb{W}^H\bs{\delta}\mb{I}_u
\label{eq.Yform_2}.
\end{align}
The proposed whitened least-square problem is:
\begin{align}
\label{eq.single_whitened1}
\ell^*&=\underset{\ell \in \mathcal{F}(t)}{\text{argmin }}||\hat{\mb{b}}-\underbrace{(\mb{S}^{-1}\mb{U}^H~|~-\mb{W}^H)}_{\mb{D}}\mb{\tilde{I}}_{E,\ell}||^2_2
\end{align} 
where $\hat{\mb{b}}=\mb{R}\mb{b}$.
The correlation of the columns of the sensing matrix $\mb{D}$ is of our interest; specifically we next investigate: 
\begin{align}
\mb{D}^H\mb{D}=\begin{pmatrix}
\mb{U}\mb{S}^{-2}\mb{U}^H&-\mb{U}\mb{S}^{-1}\mb{W}^H\\
-\mb{W}\mb{S}^{-1}\mb{U}^H&\mb{W}\mb{W}^H
\end{pmatrix}
\label{eq.corr_D}
\end{align}

The structure of \eqref{eq.corr_D} is very insightful as all the four blocks of this matrix are dependent on the singular values decomposition of $\mb{Y}_{au}\mb{Y}^{-1}_{uu}$ and, thus, this matrix 
%Recall that our definition of community is a set of nodes with high correlation among the columns related to a certain phase, 
controls the size and number of communities. 
%Since the matrix $\mb{Y}^{-1}_{uu}$ is invertible, the properties of $\mb{Y}_{au}\mb{Y}^{-1}_{uu}$ can be inferred from the properties of the rows of $\mb{Y}_{au}$, viewed as a {\it sketch} of  $\mb{Y}_{au}\mb{Y}^{-1}_{uu}$.This suggests how to explain the number and size of these communities based on the matrix $\mb{Y}_{au}$.
Since we have very few observable nodes, the rank of the matrix $\mb{Y}_{au}$ is limited by its number of rows. However, the algorithm performs best if the rows of $\mb{Y}_{au}$ are as uncorrelated from each other as possible. 
Recall, that $\mb{Y}_{au}$ is sparse and the weights (the admittance values) tend to have similar values; that means that the correlation among its rows can be largely inferred from looking at the support (non-zero elements) of the different rows of $\mb{Y}_{au}$: small overlap implies near orthogonality. Rows that have a strong overlap in the support and high correlation, on the other hand, point to parts of the graph that are direct neighbors. Bigger overlap implies that a particular community increase its size and the ambiguity to locate the actual fault location also increases.

\begin{figure}[ht]
\centering
\includegraphics[trim={0cm 0.5cm 0cm 0.5cm},width=0.3\textwidth]{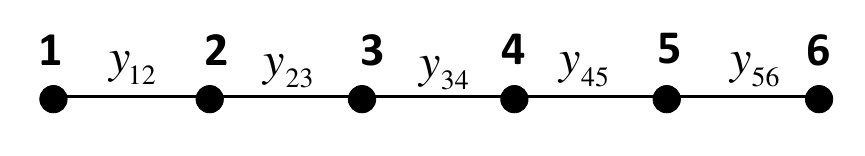}

\caption{One-Line Diagram of a Sample Radial Network}
\label{fig.ill_place}
\end{figure}

This also suggests a good sensor placement strategy, which is to have communities with the smallest sizes possible, since it leads to the least overlap among the non-zero elements of rows of $\mb{Y}_{au}$; in turn, since the admittance matrix indicates the adjacency of nodes in the grid, a good heuristic is to place sensors on parts of the grid that are more sparsely connected. 
To illustrate it, consider the one-line diagram as shown in Fig.~\ref{fig.ill_place}. 
The second block $-\mb{U}\mb{S}^{-1}\mb{W}^H$ of the correlation matrix $\mb{D}^H\mb{D}$, which corresponds to the correlation of the observable and unobservable nodes, has a sparsity pattern identical to that of of $\mb{Y}_{au}\mb{Y}^{-1}_{uu}$.
Suppose we have two $\mup$s to place in the network (which are insufficient for observability). The first requirement of a good placement is to avoid putting two $\mup$s next to each other. If that can be done, $\mb{Y}_{aa}$ is diagonal. Also, we want each observable nodes to cover as much unobservable nodes as possible, having nearly non-overlapping rows for $\mb{Y}_{au}$. We also want a nearly block diagonal structure for $\mb{Y}_{uu}$ (and therefore $\mb{Y}^{-1}_{uu}$) that mimics the non-overlapping blocks in the rows of $\mb{Y}_{au}$. 
%This means that one observable node is correlated with adjacent unobservable nodes but should not correlated with the others in the other clusters. It therefore creates a similar sparsity in the second block of $\mb{D}^H\mb{D}$ so that each observable node has the least number of correlated unobservable nodes shared with another observable node.        
Considering Fig. \ref{fig.ill_place}, if the $\mup$s are placed at bus 2 and 5, we have one such {\it good design} and the admittance matrix with blocks partitioned based on available and unavailable is as follows:
\begin{align*}
\mb{Y}=
\left(
\begin{array}{c c ;{4pt/4pt} c c |c c}
y_2&0&-y_{12}&\multicolumn{1}{c}{-y_{23}}&0&0\\ 
0&y_5&0&\multicolumn{1}{c}{0}&-y_{45}&-y_{56}\\ \hdashline[4pt/4pt]
-y_{12}&0&y_{1}&0&0&0\\
-y_{23}&0&0&y_3&-y_{34}&0\\ \cline{3-6}
0&-y_{45}&0&-y_{34}&y_4&0\\
0&-y_{56}&0&0&0&y_6
\end{array}
\right)
\end{align*}
%\begin{align*}
%\overbrace{\begin{pmatrix}
%y_1&0\\
%0&y_5
%\end{pmatrix}}^{\mb{Y}_{aa}}\rightarrow 
%&\overbrace{\begin{pmatrix}
%-y_{12}&-y_{23}&0&0\\
%0&0&-y_{45}&-y_{56}
%\end{pmatrix}}^{\mb{Y}_{au}}\\
%&\hspace{0.25cm}\underbrace{\begin{pmatrix}
%~~y_{1}&0&0&0\\
%~~0&y_3&-y_{34}&0\\
%~~0&-y_{34}&y_4&0\\
%~~0&0&0&~y_6
%\end{pmatrix}}_{\mb{Y}_{uu}}
%\end{align*}
The placement in this example is done so that all the columns of $\mb{Y}_{au}$ connect the available nodes to all the unavailable nodes. As it can be seen, the sparsity pattern of $\mb{Y}_{au}$ in this placement suggest non-overlapping support for the rows of $\mb{Y}_{au}$ to cover all the unavailable nodes.   
 What the design we just illustrated does, is making $\mb{Y}_{au}\mb{Y}^{-1}_{uu}$ already behave almost as a set of orthogonal rows, providing the best conditioning for the algorithm.  

%This would lead to a placement with sensors far enough from each other in terms of electrical distance, resulting in a good condition number, but also columns of $\mb{Y}_{au}$ which are divided in clusters that are naturally locally close on the topology, allowing to discriminate the faults with the highest possible resolution, for a given number of sensors.
%
%
%This interpretation gives us an insight about the reason of having these communities of neighboring nodes correlated with each other and provide an intuition about a good versus bad sensor placement strategy. We leave the formal derivation of placement strategy for future work.    

\section{Case Study}\label{sec.numerical}
\begin{figure}[ht]
\centering
\includegraphics[trim={0cm 0.5cm 0cm 0.7cm},width=0.45\textwidth]{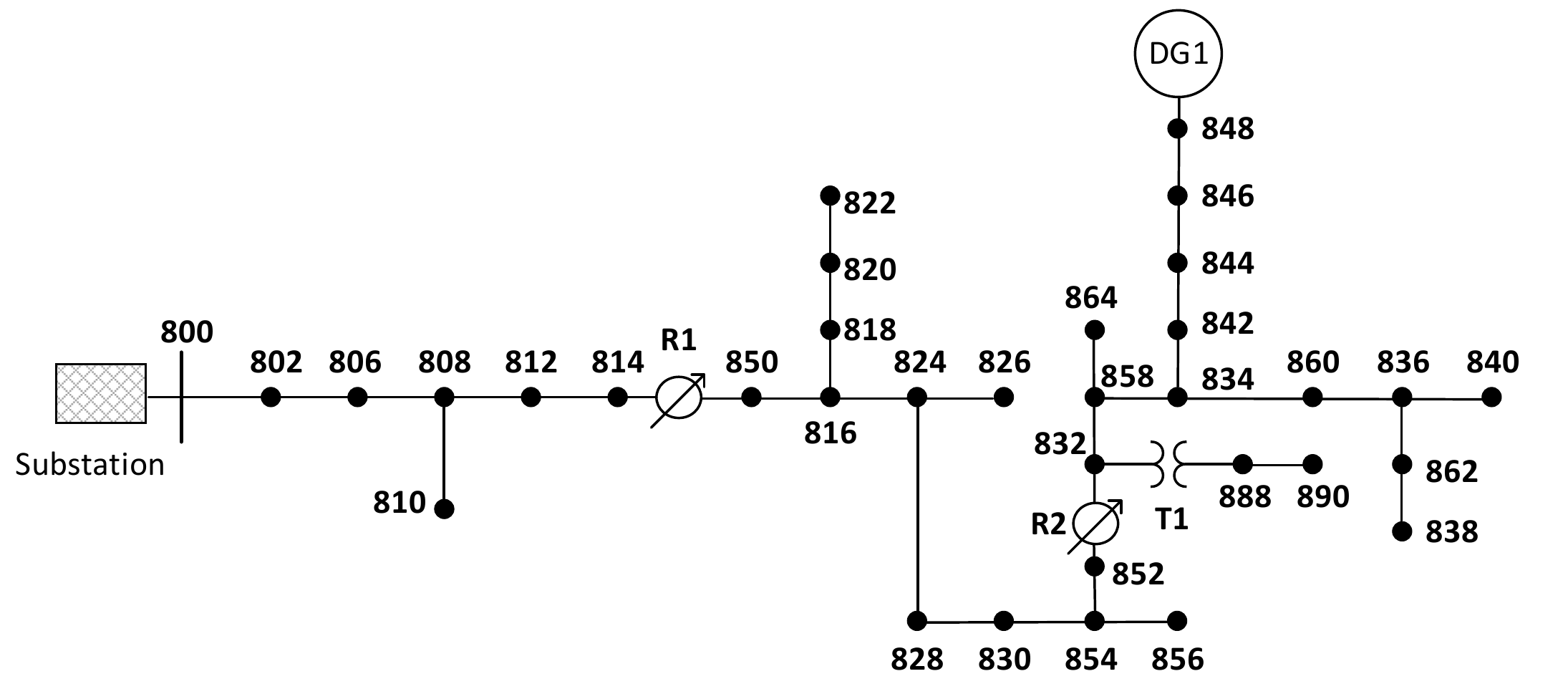}
\caption{One-Line Diagram of IEEE-34 Test Case with Added Generator.}
\label{fig.ieee34}
\end{figure}

As a case study, we use IEEE-34 bus test case~\cite{kersting2001radial}, where a 100 kW generator is added at bus 848. This radial grid is an unbalanced grid with untransposed lines and one phase laterals. The one-line diagram of the test case is shown in Fig.~\ref{fig.ieee34}.
OpenDSS software is used to perform the analysis here~\cite{dugan2012reference}. A snapshot of power flow is solved to get the data before fault and then a fault is introduced in the dynamic mode to represent the behavior of the grid after a fault occurs. Note that the tap changers usually have a delay for 15-30 seconds in their settings to respond to a change so the voltage and current data should be recorded before the tap values change so that the required bus impedance matrix in the formulation stays the same before and after the fault. 

Assume that 5 $\mup$s are available placed on the three-phase nodes to provide the three phase voltage and current flowing in the lines incident to the bus, to which they are connected. The nodes with $\mup$ are listed in Table.~\ref{tab.locations}. 
%Table.~\ref{tab.locations} provides the locations of the $\mup$s on the grid. 
\begin{table}[ht]
\centering
\caption{}
\begin{tabular}{|c|c|c|}
\hline
Test Case&\#$\mup$s&Location\\ \hline
IEEE-34&5&800-830-848-832-862 \\ \hline
\end{tabular}
\label{tab.locations}
\end{table} 
As it can be seen, two of the $\mup$s are located next to source buses that contribute to the fault current in order to obtain an approximation for the fault current vector $\mb{I}_G$.

\subsection{Limitation Evaluation of the Metric in \eqref{eq.single_form1}}
In this section, through some numerical analysis, we show the limitations of the proposed metric in \eqref{eq.single_form1}, corroborating our early discussions.

\subsubsection*{Scenario 1}
\begin{figure}[ht]
\centering
\includegraphics[trim={0cm 0.1cm 0cm 1cm},width=0.5\textwidth]{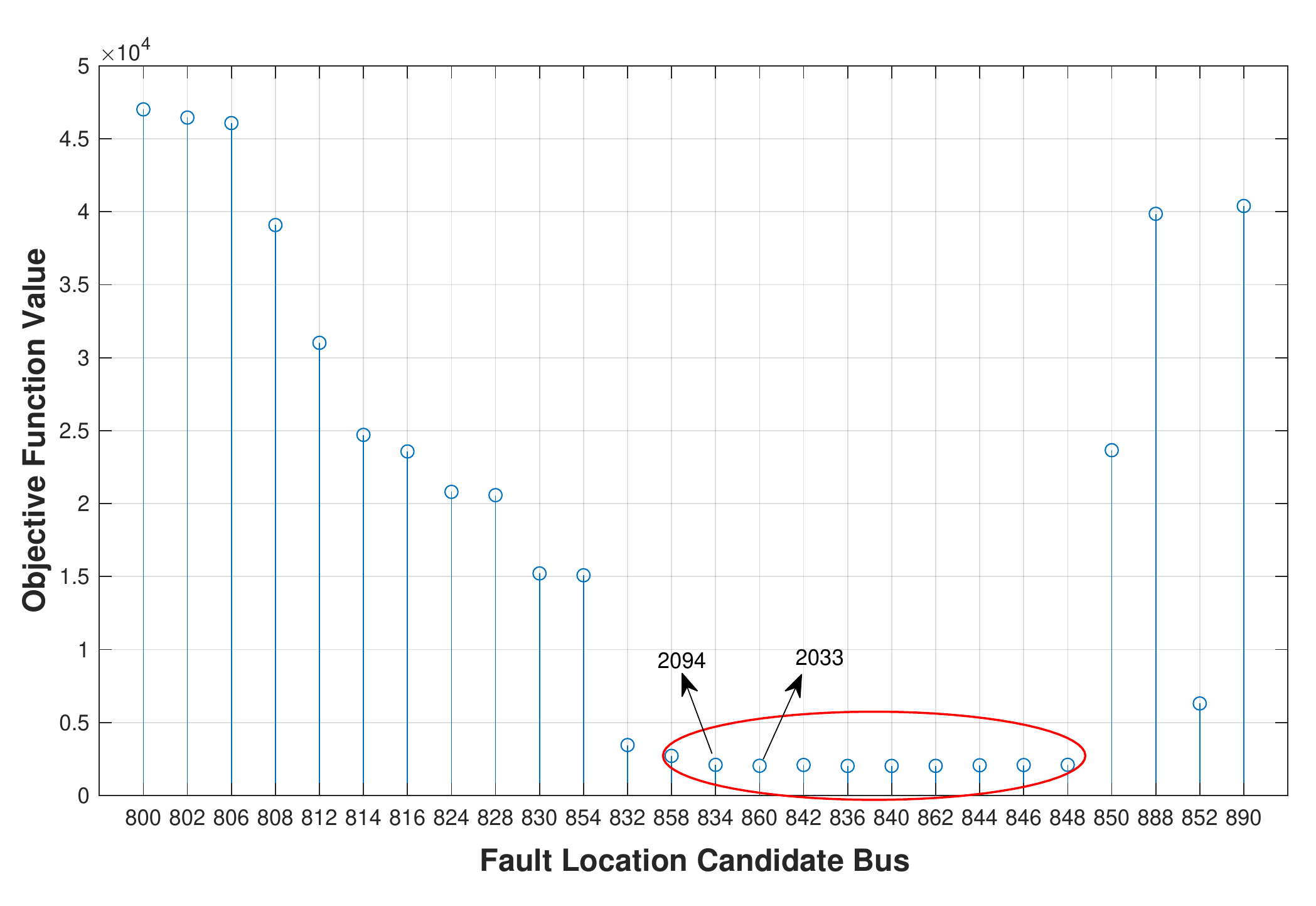}
\caption{Objective Function Value of \eqref{eq.single_form1} for Three-Phase Fault-Bus 834 .}
\label{fig.failed_1}
\end{figure}

In this scenario, a three-phase fault is introduced at bus 25 and the fault current is approximated using the data from sensors at buses 800 and 848. Fig.~\ref{fig.failed_1} shows the value of the squared norm of the objective function in \eqref{eq.single_form1} for the candidate locations, where a three-phase fault could potentially happen. 
As it can be observed, the value of the objective function is very close for a set of neighboring nodes that can be a candidate for the fault. The algorithm can simply pick a wrong location if the data is corrupted with noise. Even with clean data, it can be seen that the objective function is returning a smaller value for bus 860 as opposed to the value for bus 834 (actual location of the fault). That can be attributed to the fact that we only have access to approximate fault current and also to the fact that part of the grid is unobservable.  
\subsubsection*{Scenario 2}
\begin{figure}[b]
\centering
\includegraphics[trim={0cm 0.5cm 0cm 1.5cm},width=0.82\linewidth]{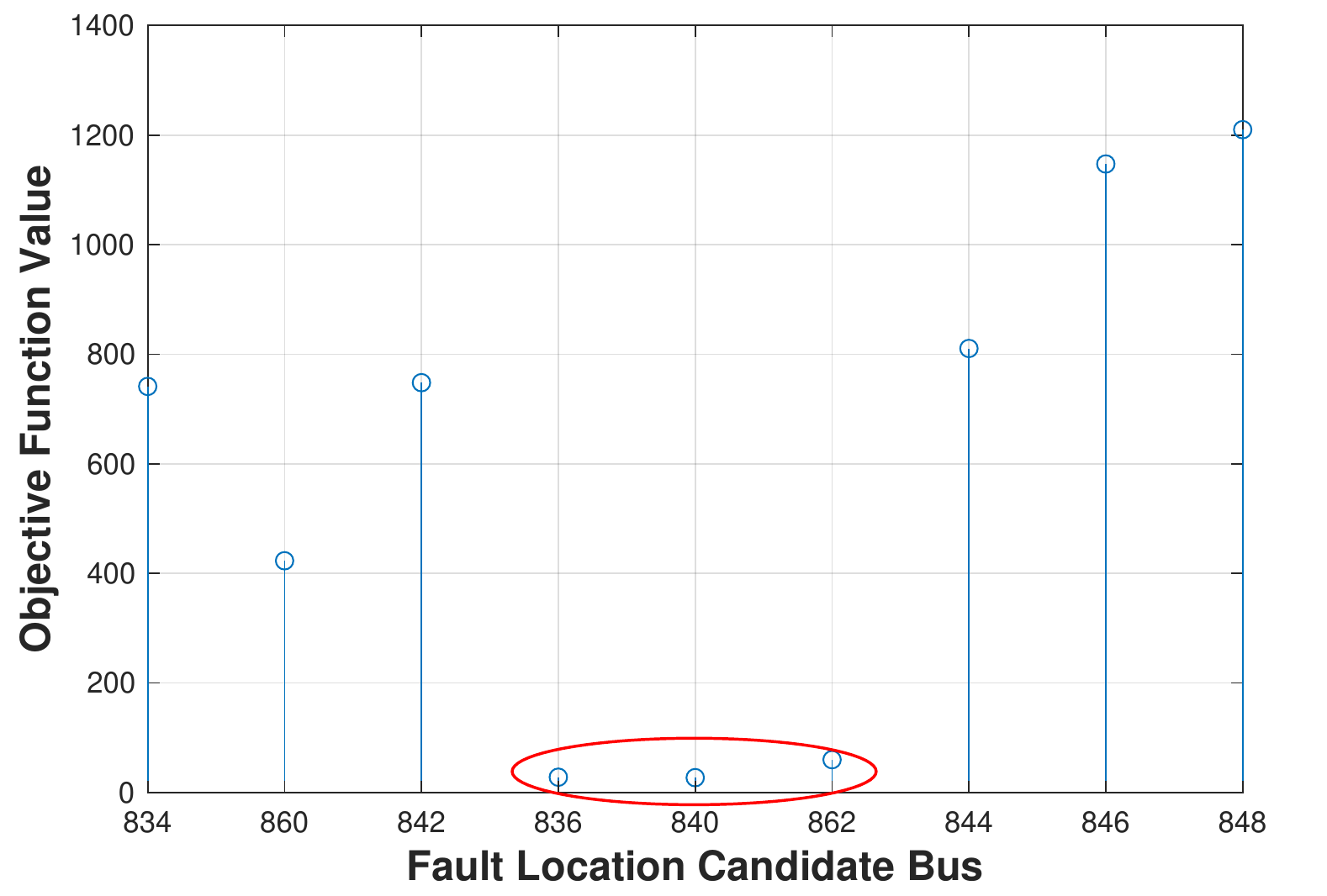}
\caption{Objective Function Value of \eqref{eq.single_form1} for Three-Phase Fault-Bus 836 .}
\label{fig.failed_2}
\end{figure}

In this scenario, we assume that the exact value of the fault current is somehow given to us. The reason for this assumption is to show that the approximate fault current is not only the cause for the aforementioned ambiguity and, in fact, the similarity of the corresponding columns of $\mb{Z}_a$ is the root cause of this ambiguity. In this scenario, a three-phase fault is introduced at bus 836. The result of the metric for this case is given in Fig.~\ref{fig.failed_2}.
It is clear from the results that there is an inherent ambiguity in locating the fault in this case even when the exact fault current is available: due to the correlation of the columns in $\mb{Z}_a$ buses 840, and 862 are prone to be mistaken with bus 836 as the fault location.

\subsection{Test-Case Community of Nodes}
In this part, the correlation of the columns of the matrix pre-multiplied by $\mb{\tilde{I}}_{E,\ell}$ in \eqref{eq.single_form1} and \eqref{eq.single_whitened1} is shown. We use the following definition to look at the normalized value of the unsigned correlation of the columns. 
\begin{definition}
The absolute value for the correlation coefficients of the columns of a matrix $\mb{X}=[\mb{x}_1,\mb{x}_2,\hdots]$ that is used is defined as follows:
\begin{align}
[\mb{C}]_{m,n}=\frac{|\mb{x}_m^H \mb{x}_n|}{||\mb{x}_m||~||\mb{x}_n||}
\label{eq.correlation}
\end{align} 
\end{definition}   

\begin{figure}[ht]
\centering
\includegraphics[trim={0cm 0.5cm 0cm 0.5cm},width=0.5\textwidth]{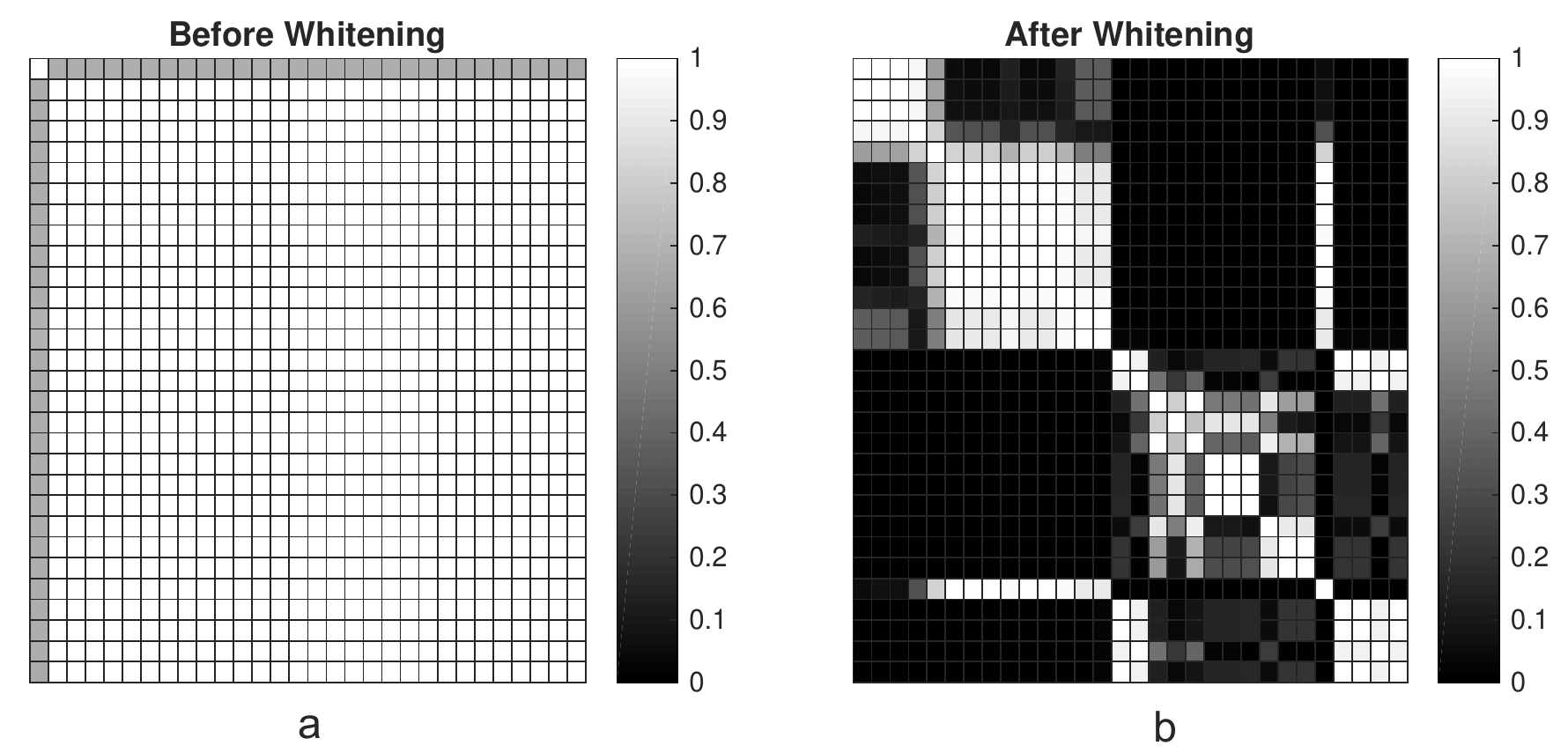}
\caption{Correlation of Columns related to Phase-A of a) $\mb{Z}_a$ and b) $\mb{D}$.}
\label{fig.heatmap}
\end{figure}

Fig.~\ref{fig.heatmap} shows the correlation of the columns of $\mb{Z}_a$ (on the left) and $\mb{D}$ (on the right) corresponding to phase-A that is produced using the definition\footnote{Note that the correlation between the columns corresponding to phase $i~(i=a,b,c)$ and $j~(j\neq i)$ is not important since for a faulty phase $i$, the indices corresponding to phase $j$ are not candidates. 
%For example, when a three phase fault occurs, we want to put the first element of fault current vector at locations corresponding to phase $a$ so it is important how these columns corresponding to phase $a$ form communities no matter what their correlations are with respect to the columns corresponding to phase $b$ or $c$.
}
given in \eqref{eq.correlation} based on the placement in Table.~\ref{tab.locations}.  
  
In this figure,\footnote{The ordering of the nodes have changed here to put the neighboring nodes as close as possible to each other to better visualize the communities, whereas the actual matrix is separated as blocks of available and unavailable nodes.} the correlated columns of $\mb{D}$ in Fig.~\ref{fig.heatmap}b in our proposed method reveal clusters of nodes that a fault can be localized up to their level, whereas these clusters do not appear in the correlation of the columns of the sensing matrix $\mb{Z}_a$ in \eqref{eq.single_form1} as shown in Fig.~\ref{fig.heatmap}(a) so the performance of the fault localizer cannot be investigated in this method. 
 
\begin{figure}[ht]
\centering
\includegraphics[trim={0cm 1cm 0cm 0.5cm},width=0.5\textwidth]{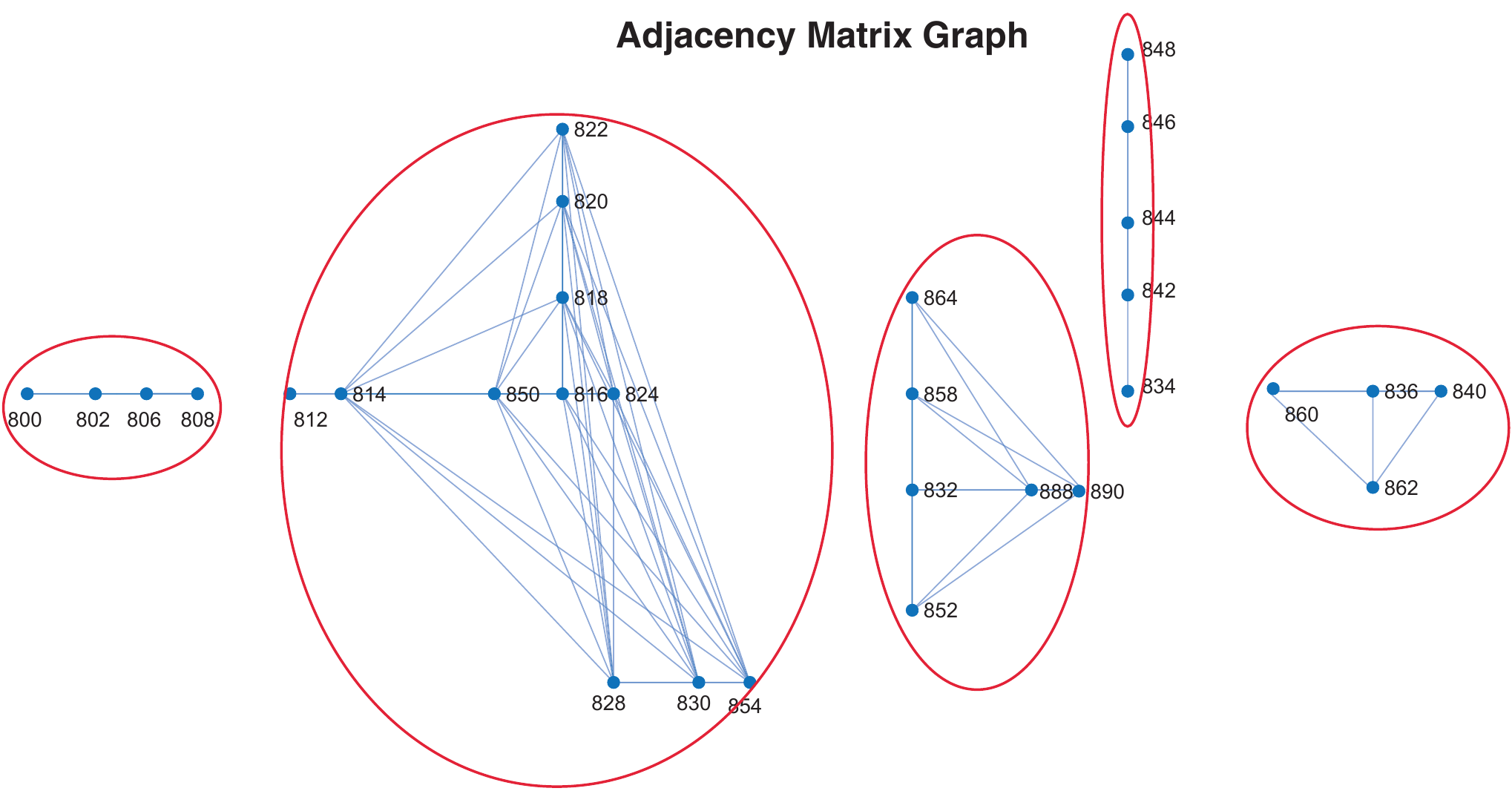}
\caption{Adjacency Matrix Graph for Correlation Coefficients of Columns of $\mb{D}$ with $\tau=0.814$.}
\label{fig.graph}
\end{figure}

To better understand the location of highly correlated nodes with respect to each other, we first choose a threshold of $\tau$ on the correlation coefficient and build an adjacency matrix $\mb{A}$ as follows:
\begin{align}
[\mb{A}]_{m,n}=
\begin{cases}
1& \text{if }[\mb{C}]_{m,n}\geq \tau, m \neq n \\
0& \text{else}
\end{cases}
\end{align}
Fig.~\ref{fig.graph} shows a graph corresponding to the adjacency matrix $\mb{A}$, built using the correlation coefficients of columns of $\mb{D}$ and overlaid on the IEEE-34 test case topology.  
The heat-map for the correlation coefficients of the columns corresponding to phase-B and phase-C in the matrix $\mb{D}$ follow a similar pattern as in Fig.~\ref{fig.heatmap}(b).
As expected based on our analysis in Section \ref{sec.communities}, the nodes with high correlation are those that are located in a neighborhood of each other. The fault location in the presented approach can locate the fault up to the resolution of these communities, which can be interpreted as a low-resolution representation of the graph. 

%\footnote{The heat-maps are generated by keeping the correlation coefficients above $\tau$ as it is and put other coefficients equal to zero} of the columns corresponding to phase-B and phase-C in the matrix $\mb{D}$ follow a similar pattern, as one can observe in Fig.~\ref{fig.heatmap_B_C}.   
% \begin{figure}[ht!]
%\centering
%\includegraphics[width=0.5\textwidth]{figures/heatmap_B_C.pdf}
%\caption{Thresholded Correlation Coefficients of Columns related to Phase-B and Phase-C of $\mb{D}$ with $\tau=0.814$.}
%\label{fig.heatmap_B_C}
%\end{figure} 

\begin{figure}[ht!]
\centering
\includegraphics[trim={0cm 0.5cm 0cm 0.5cm},width=0.35\textwidth]{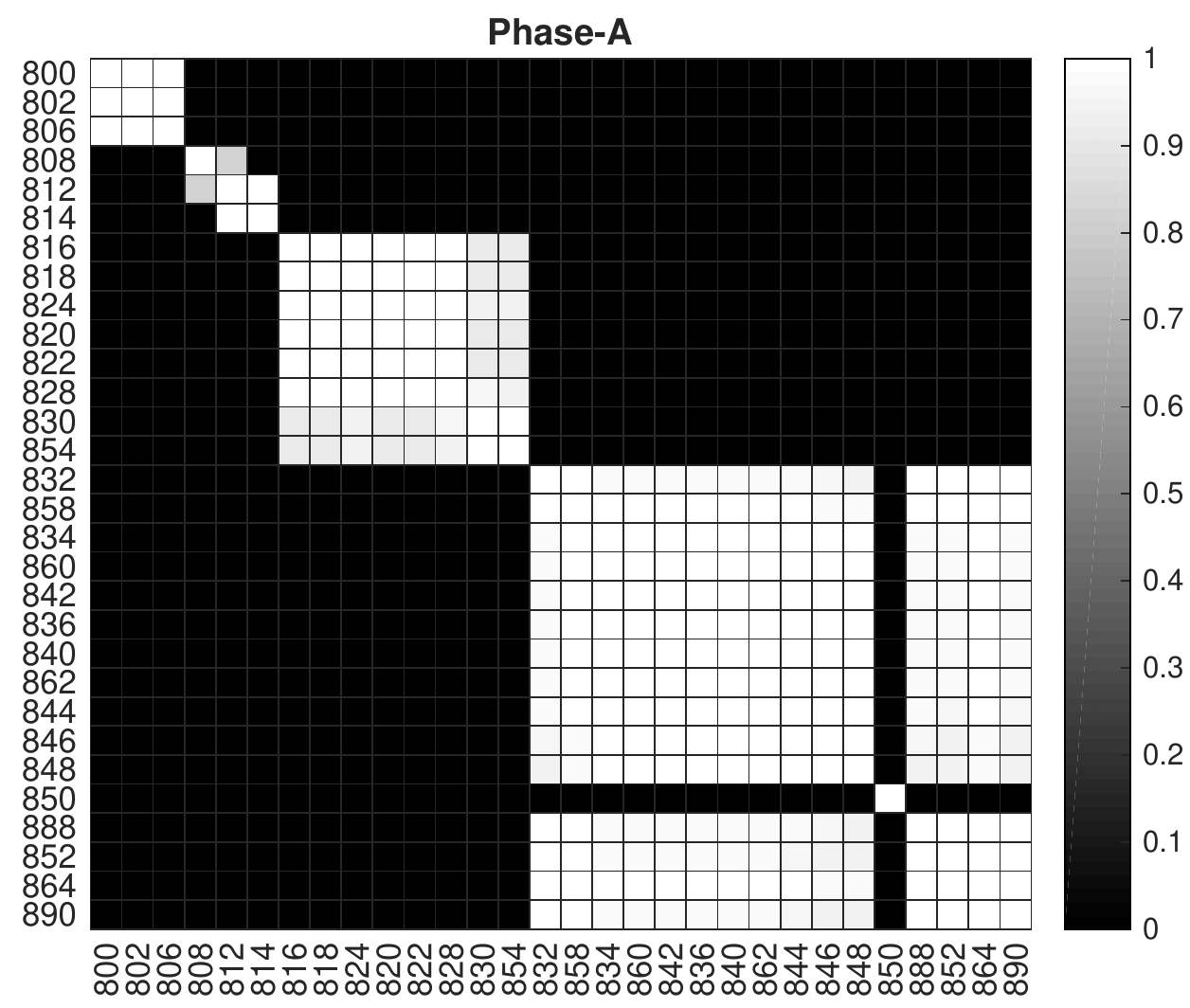}
\caption{Thresholded Correlation Coefficients of Columns related to Phase-A $\mb{D}$ with $\tau=0.814$.}
\label{fig.BadLoc_heatmap}
\end{figure}

It should be noted that the communities that emerge are dependent on the locations of the sensors. The placement based on Table~\ref{tab.locations} has been done leveraging the heuristic discussed at the end of Section \ref{sec.communities}. To show how a bad placement change the detected communities, we place the sensors according to Table~\ref{tab.BadLoc}.
\begin{table}[ht]
\centering
\caption{}
\begin{tabular}{|c|c|c|}
\hline
Test Case&\#$\mup$s&Location\\ \hline
IEEE-34&5&800-814-816-848-850 \\ \hline
\end{tabular}
\label{tab.BadLoc}
\end{table}  
The thresholded heat-map corresponding to phase-A is shown in Fig.~\ref{fig.BadLoc_heatmap}.
It can be seen that since the $\mup$s are more condensed on the left side of the grid, the ambiguity in detecting faults among nodes where there is no sensor increases. On a separate test, we saw that the placement according to Table~\ref{tab.locations} results in a $\mb{Y}_{au}$ that is of full row rank, whereas the placement according to Table~\ref{tab.BadLoc} makes $\mb{Y}_{au}$ extremely ill-conditioned.

\subsection{Fault Location Using Metric in \eqref{eq.single_whitened1}}
In this part, we introduce different types of faults and use the metric in \eqref{eq.single_whitened1} to locate the fault assuming that the sensors are placed according to Table~\ref{tab.locations}. Even though the new formulation improves the fault location algorithm, there is still an ambiguity involved, and localization is reliable at the community level but not at the bus level. 
%In other words, the localization can be achieved at low-resolution up to the community level. 

Table.~\ref{tab.faults} summarizes the results for three different types of faults. 
\begin{table}[ht]
\centering
\caption{Identified Fault Locations with Metric in \eqref{eq.single_whitened1}}
\begin{tabular}{|c|c|p{40mm}|}
\hline 
Fault Type&Exact Fault Location&Locations with Close Objective \newline Values in \eqref{eq.single_whitened1}\\ \hline \hline
LLL&816&814-816-850\\ \hline
A-G&822-A&814-A,816-A,818-A,820-A,\newline 822-A,850-A\\ \hline
BC-G&852-B-C&832-B-C,852-B-C\\
\hline
\end{tabular}
\label{tab.faults}
\end{table}   
The first column in this table is the type of the introduced fault and the second column indicates the exact location, where the fault is introduced. In the third column, a list of candidate locations, for which the objective function values in \eqref{eq.single_whitened1} are close to each other is listed. These locations can be mistaken with the exact location due to some noisy measurements or poor fault current approximation, or not having enough sensors compared to the size of the network. The comparison of the results in this table with the communities detected in Fig.~\ref{fig.graph} confirms the claim that errors are concentrated within the same community where the exact fault location exists.   

\section{Future Work and Concluding Remarks}
We investigated the performance of distribution line fault localization. Our results showed that using our proposed method, a fault is identifiable up to a level of ``community'' of neighboring nodes, which is referred to as ``low-resolution'' fault localization. We also discussed the effect of sensor placements on the size and the number of these communities and provided algebraic interpretation for a good versus a bad sensor placement. Our future work includes a mathematical way to formulate an optimal sensor placement when a limited number of $\mup$s are available in order to have the highest fault localization resolution.    
                            
%\section*{Acknowledgment}
%Acknowledgment: This material is based upon work supported by the Department of Energy under Award Number DE-OE0000780.      
%
%\section*{Disclaimer} This paper was prepared as an account of work sponsored by an agency of the United States Government. Neither the United States Government nor any agency thereof, nor any of their employees, makes any warranty, express or implied, or assumes any legal liability or responsibility for the accuracy, completeness, or usefulness of any information, apparatus, product, or process disclosed, or represents that its use would not infringe privately owned rights. Reference herein to any specific commercial product, process, or service by trade name, trademark, manufacturer, or otherwise does not necessarily constitute or imply its endorsement, recommendation, or favoring by the United States Government or any agency thereof. The views and opinions of authors expressed herein do not necessarily state or reflect those of the United States Government or any agency thereof.
                            
\bibliography{bib}
\vspace{-0.1cm}
\bibliographystyle{IEEEtran}

\end{document}